\documentclass[aps,prl,twocolumn,superscriptaddress]{revtex4-1}
\usepackage[utf8]{inputenc}
\usepackage{amsmath}
\usepackage{amssymb}
\usepackage{epsfig}
\usepackage{graphicx,url}
\usepackage{bm}
\usepackage{mathrsfs}
\usepackage{color}

\begin{document}

\title{On the ``generalized Generalized Langevin Equation'' }

\author{Hugues Meyer}
\affiliation{\it Physikalisches Institut, Albert-Ludwigs-Universit\"{a}t,  79104 Freiburg, Germany}
\affiliation{Research Unit in Engineering Science, Universit\'{e} du Luxembourg,\\  L-4364 Esch-sur-Alzette, Luxembourg}
\author{Thomas Voigtmann}
\affiliation{Institut f\"{u}r Materialphysik im Weltraum,
Deutsches Zentrum f\"{u}r Luft- und Raumfahrt (DLR),\\ 51170 K\"{o}ln, Germany}
\affiliation{Department of Physics, Heinrich Heine University, Universit\"{a}tsstra\ss e 1, 40225 D\"{u}sseldorf, Germany}
\author{Tanja Schilling}
\affiliation{\it Physikalisches Institut, Albert-Ludwigs-Universit\"{a}t,  79104 Freiburg, Germany}

\date{\today}

\begin{abstract}
In molecular dynamics simulations and single molecule experiments, observables are usually measured along dynamic trajectories and then averaged over an ensemble (``bundle'') of trajectories. Under stationary conditions, the time-evolution of such averages is described by the generalized Langevin equation. In contrast, if the dynamics is not stationary, it is not a priori clear which form the equation of motion for an averaged observable has. We employ the formalism of time-dependent projection operator techniques to derive the equation of motion for a non-equilibrium trajectory-averaged observable as well as for its non-stationary auto-correlation function. The equation is similar in structure to the generalized Langevin equation, but exhibits a time-dependent memory kernel as well as a fluctuating force that implicitly depends on the initial conditions of the process. We also derive a relation between this memory kernel and the autocorrelation function of the fluctuating force that has a structure similar to a fluctuation-dissipation relation. In addition, we show how the choice of the projection operator allows to relate the Taylor expansion of the memory kernel to data that is accessible in MD simulations and experiments, thus allowing to construct the equation of motion. As a numerical example, the procedure is applied to Brownian motion initialized in non-equilibrium conditions, and is shown to be consistent with direct measurements from simulations.
\end{abstract}

\maketitle

\subsection{Introduction}
Consider a system of $N$ classical particles that evolve according to Hamilton's equations of motion, and a phase space variable $A_{t} \equiv A (\mathbf{q}^{N}(t),\mathbf{p}^{N}(t))$, where $\mathbf{q}^{N}(t)$ and $\mathbf{p}^{N}(t)$ are the positions and momenta of the particles. In several sub-fields of statistical physics quantities such as $\langle A_tA_{t'}^*\rangle_{\rm Trajectories}$ are studied, i.e.~averages of auto-correlations in $A$ taken over bundles of trajectories. (If the reader wonders why we introduce the auto-correlation rather than the simpler expression $\langle A_t \rangle_{\rm Trajectories}$, which is of equal practical importance, we suggest to skip ahead and compare eqns.~(\ref{EOM_A_full}) and (\ref{EOM_C}).~ Note that the latter expression is easier to analyze, because the fluctuating force averages out.) 

One context, in which this type of quantity is relevant, is the field of molecular dynamics simulation of microscopic systems with the aim of constructing coarse-grained models: for instance, the system could be a polymer melt and the aim could be to develop a rheological model \cite{padding:11}; or the system could be a biomolecule, which a researcher might simulate with classical atomistic force-fields and monitor the collective motion of specific groups of atoms in order to deduce a simplified model of a biological mechanism \cite{berendsen:07, kmiecik:16}. A different and equally important context is the experimental study of non-equilibrium work-relations in single molecule experiments \cite{harris:04, hoffmann:12}, where $A_t$ would be e.g.~the extension of a piece of DNA or of a protein under an applied force. 

In any of these contexts, it is useful to have general information on the properties of the equation of motion that governs $\langle A_tA_{t'}^*\rangle_{\rm Trajectories}$. Therefore, in this article we discuss the form of the equation of motion in the general (i.e.~the non-stationary) case.

The problem of integrating out a large number of degrees of freedom of a thermodynamic system can be treated by several approaches. Most widely used is the framework of stochastic equations of motion, which has been introduced by Langevin \cite{langevin:1908} with his description of Brownian motion. In the 1960's a formalism was developed by Zwanzig \cite{zwanzig:1961} and Mori \cite{mori:1965} to account for memory effects in non-trivial systems. This formalism is based on the definition of projection operators that are aimed at integrating out the fast dynamics of a process in order to study slow variables. 

Since then, several approaches were introduced to extend the formalism to non-stationary dynamics \cite{kawasaki:1973,willis:1974,furukawa:1979,grabert:1982,linden:1998,fuchs:2009}. We will, in the following base our arguments on Grabert's approach \cite{grabert:1982} and introduce a new projection operator, that is particularly suited to study variables of the type $\langle A_tA^*_{t'}\rangle$.  (To shorten the notation, we have dropped the subscript ``Trajectories''. Unless stated otherwise, in the following all averages are meant as trajectory averages.) Note that the relations discussed here do not require time-scale separation, i.e.\ the averaged observable can be any variable of interest, regardless of whether it is 'slow' or 'fast', 'relevant' or not.

\subsection{The stationary case}
Before we address the non-stationary case, let us briefly recall the structure of the equations in equilibrium or other stationary situations. The evolution of the the variable $A_{t}$ is governed by the equation
\begin{equation}
\label{simpleGLE1}
\frac{dA_{t}}{dt} = \omega A_{t} + \int_{0}^{t}K(t-\tau)A_{\tau}d\tau + \eta(t)
\end{equation}
where $\omega$ is a drift coefficient, $K(t)$ is a so-called memory kernel, and $\eta(t)$ is a so-called fluctuating force \cite{hansen:1990}. (We denote the time-dependence of operators by a subscript in contrast to the time-dependence of functions, which we denote in brackets). This equation has the form of a generalized Langevin equation (GLE). The corresponding auto-correlation function $C(t) = \left<A_{t}A_{0}^{*}\right> = \left<A_{t+t'}A_{t'}^{*}\right>$ evolves according to
\begin{equation}
\label{simpleGLE}
\frac{dC(t)}{dt} = \omega C(t) + \int_{0}^{t}K(t-\tau)C(\tau)d\tau 
\end{equation}
 The dependence of $K$ on $t-\tau$ in the integral is convenient because the convolution theorem can be used to Laplace transform equation (\ref{simpleGLE}) \cite{mokshin:2005}.
Thus if the autocorrelation function $C(t)$ and the drift coefficient $\alpha$ can be obtained in an experiment or simulation, the stationary memory kernel can be constructed. Note that under non-stationary conditions this property does not hold, i.e.~one can in general not obtain the memory kernel by a simple Laplace transform of the observed dynamics of a coarse-grained variable.

\subsection{Time-dependent projection operators: a brief reminder}

To introduce the non-stationary case, we briefly recall Grabert's approach \cite{grabert:1982}. Consider the time-evolution of the dynamical variable $A_{t}$
\begin{equation}
\label{liouville_eom}
\frac{dA_{t}}{dt} = i\mathcal{L}A_{t}
\end{equation}
where $i\mathcal{L}$ is the propagator, e.g.~the Liouvillian operator in the case of Hamiltonian dynamics. (Note that the following arguments are not restricted to Hamiltonian dynamics. They hold for any dynamics that can be described by an equation of the form of eqn.~(\ref{liouville_eom}). Thus our conclusions also apply, in particular, to simulations with a Nos{\'e}-Hoover thermostat.) Equation (\ref{liouville_eom}) can be formally integrated and then again differentiated with respect to time to yield
\begin{equation}
\label{dadt}
\frac{d A_{t}}{dt} = e^{i\mathcal{L}t}i\mathcal{L}A_{0}
\end{equation}

In the equilibirum Mori-Zwanzig formalism, one usually defines a stationary projection operator which is used to split the dynamics into a parallel (slow) part and on orthogonal (fast) part. When extending the formalism to non-stationary processes, Grabert introduced the time dependence directly in the projection operator. Assume that one can define a time-dependent operator $P_{t}$ that acts on phase space variables and that fullfills
\begin{equation}
\label{prop_Pt}
P_{t'}P_{t} = P_{t}
\end{equation}
for all times $t$ and $t'$. Note that if $t'=t$, one obtains $P_{t}^{2} = P_{t}$, i.e.~this operator is a projector. In fact, it indicates that $P_{t}$ projects onto a fixed subspace for all $t$, but the orientation of the projection might change with $t$. This implies that the projection of a vector with respect to a certain $t$ is part of the fixed subspace and thus remains constant once projected with respect to another time $t'$. We can then take the derivatives with respect to either $t$ or $t'$, and then take the limit $t' \rightarrow t$ to find 
\begin{align}
{P}_{t}\dot{{P}}_{t} &= \dot{{P}}_{t} \label{prop_Pdot1} \\
\dot{{P}}_{t}{P}_{t} &= 0
\end{align}
Thus, we can write
\begin{equation}
\label{prop_Pdot}
P_{t}\dot{P}_{t}(1-P_{t}) = \dot{P}_{t}
\end{equation}
Now we define the operator $Z_{t}=e^{i\mathcal{L}t}(1-P_{t})$. Its time-derivative is
\begin{equation}
\dot{Z}_{t} = e^{i\mathcal{L}t}i\mathcal{L}(1-P_{t})  -e^{i\mathcal{L}t}\dot{P}_{t}  
\end{equation}
Inserting $1 = \mathcal{P}_{t} + (1-\mathcal{P}_{t})$ in the first term and using eqn.~(\ref{prop_Pdot}) yields
\begin{equation}
\label{diffeq_Z}
\dot{Z}_{t} = Z_{t}i\mathcal{L}(1-P_{t}) + e^{i\mathcal{L}t}P_{t}(i\mathcal{L}-\dot{P}_{t})(1-P_{t})
\end{equation}
This differential equation for $Z_{t}$  can be solved using time-ordered exponentials,
\begin{align}
\label{solve_Z}
Z_{t} &= e^{i\mathcal{L}s}\left[1-P_{s} \right]G_{s,t} \nonumber \\
&+ \int_{s}^{t}{d\tau e^{i\mathcal{L}\tau}  \left[i\mathcal{L}-\dot{P}_{\tau}\right]\left[1-P_{\tau} \right] G_{\tau,t} }
\end{align}
which is valid for any reference time $s\le t$. In this expression $G_{\tau,t}$ is the negatively time-ordered exponential operator, i.e. the unique solution of the differential equation $dY_{\tau,t}/dt = Y_{\tau,t}i\mathcal{L}(1-P_{t})$ with initial condition $Y_{t,t}=1$. It can be written as 
\begin{widetext}
\begin{equation}
\label{time_order}
G_{\tau,t}=\exp_-[\int_{\tau}^tdt'i\mathcal L(1-P_{t'})] \equiv 1 + \sum_{n=1}^{\infty} \int_{\tau}^{t} dt_{1} \int_{\tau}^{t_{1}}dt_{2} \cdots \int_{\tau}^{t_{n-1}}dt_{n} i\mathcal{L}(1-P_{t_{n}})\cdots i\mathcal{L}(1-P_{t_{2}})i\mathcal{L}(1-P_{t_{1}})
\end{equation}
\end{widetext}
We now split the propagator $e^{i\mathcal{L}t}$ into $e^{i\mathcal{L}t}P_{t} + e^{i\mathcal{L}t}(1-P_{t})$, which allows us to rewrite eqn.~(\ref{dadt})
\begin{equation}
\frac{dA_{t}}{d t} = e^{i\mathcal{L}t}P_{t}i\mathcal{L}A_{0} + Z_{t}i\mathcal{L}A_{0}
\end{equation}
By making use of eqn.~(\ref{solve_Z}), we obtain the following equation of motion for $A_t$
\begin{align}
\label{EOM_A}
\frac{dA_{t}}{dt} =& e^{i\mathcal{L}t}P_{t}i\mathcal{L}A_{0} \nonumber \\
&+ \int_{s}^{t}{d\tau e^{i\mathcal{L}\tau}P_{\tau} \left[i\mathcal{L} - \dot{P}_{\tau} \right]\left[1-P_{\tau}\right]G_{\tau,t}i\mathcal{L}A_{0}} \nonumber \\
&+ e^{i\mathcal{L}s}\left[1-P_{s}\right]G_{s,t}i\mathcal{L}A_{0}
\end{align}
This equation is valid for any projection operator as long as it satisfies the identity (\ref{prop_Pt}). Here we close the reminder of Grabert's work and come to new aspects. 

\subsection{A time-dependent projection operator for bundles of trajectories}

We specify a particular projector $P^b_{t}$ by defining its action on any function of phase space $F$. As we intend to apply the technique to bundles of trajectories -- be they generated by a set of experiments or of molecular dynamics simulations -- we introduce a definition that is natural in this context (this is where our work differs from previous work on time-dependent projection operators). In an MD simulation, one typically initializes a bundle of trajectories at a given distribution of points in phase space, $\rho(\mathbf{\Omega}_{0})$, then one numerically propagates them and computes the variable of interest $A$ on each trajectory at certain times $t$. Finally one takes the average of $A_{t}A^{*}_{0}$ over all simulated trajectories. (A set of experiments is carried out and analyzed in exactly the same way, although $\rho(\mathbf{\Omega}_{0})$ can usually not be prescribed.) 

We thus define a time-dependent projection operator $P^{\rm b}_{t}$ by its action on a dynamical variable $F$. 
\begin{equation}
\label{projector}
P^{\rm b}_{t}F := \frac{\left\langle A^{*}_{t}F_{t} \right\rangle}{\left\langle |A_{t}|^{2} \right\rangle} A_{0}
\end{equation}
where the brackets mean an average over all possible trajectories (indicated by the superscript b for ``bundle'') starting from a well-defined distribution of initial configurations, i.e.
\begin{equation}
\label{traj_ave}
\left\langle X_{t} \right\rangle = \int d\mathbf{\Omega}_{0} \rho(\mathbf{\Omega}_{0})   e^{i\mathcal{L}t}\,X(\mathbf{\Omega}_{0})
\end{equation}
This average is also used to define the
correlation function between two dynamical variables $X$ and $Y$ by
\begin{equation}
\label{traj_ave2}
\left\langle X_{t} Y_{t'} \right\rangle = \int d\mathbf{\Omega}_{0} \rho(\mathbf{\Omega}_{0})\left[ e^{i\mathcal{L}t'}Y(\mathbf{\Omega}_{0})\right] \left[e^{i\mathcal{L}t}X(\mathbf{\Omega}_{0})\right]
\end{equation}
Here, the exponential operators are understood to act only on arguments
inside the enclosing square brackets. (In later expressions, operators are
understood as acting on all arguments that appear to their right inside
the angular brackets that denote subsequent averaging.)

\begin{figure}
	\begin{center}
		\includegraphics[width=.99\linewidth]{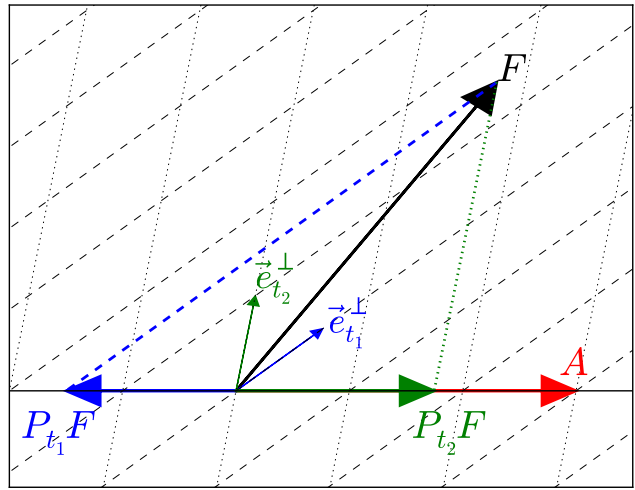}
	\end{center}
	\caption{Schematic visualization of the time-dependent projector $P^{\rm b}_{t}$. One projects a vector $F$ onto a fixed vector $A$, but the basis used to project changes with time. Equation (\ref{prop_Pt}) is illustrated by this drawing.}
	\label{proj}
\end{figure}

Compare $P^{\rm b}_{t}$ to the projection operator which is used in the stationary Mori-Zwanzig formalism, $P^{\rm MZ}F = \left\langle AF \right\rangle_{\rm eq}\left\langle A^{2} \right\rangle_{\rm eq}^{-1}A_{0}$, where $\left\langle \cdots \right\rangle_{\rm eq}$ stands for the equilibrium ensemble average. In the case of equilibrium processes, our projector collapses with $P^{\rm MZ}$. Note that the property (\ref{prop_Pt}) is satisfied. $P^{\rm b}_{t}$ can be seen as an operator that projects onto the fixed vector $A$ but whose orientation of projection evolves with time (see fig.~(\ref{proj})). 
We now apply definition (\ref{projector}) to eqn.~(\ref{EOM_A}). To simplify the resulting equation, we introduce the following notation :
\begin{equation}
\label{omega_def}
\omega_{n}(t) = \frac{\left\langle A^{*}_{t}A^{(n)}_{t} \right\rangle}{\left\langle |A_{t}|^{2} \right\rangle}
\end{equation}
where $A^{(n)}_{t}$ stands for the $n$-th time derivative of $A_{t}$, i.e. $A^{(n)}_{t} \equiv (i\mathcal{L})^{n}A_{t}$. Using this notation, the first term of eqn.~(\ref{EOM_A}) becomes $\omega_{1}(t) A_{t}$. The second term requires more attention. In eqn.~(\ref{EOM_A}) the integrand has on its left side the operator $e^{i\mathcal{L}\tau}P_{\tau}$, which implies that it is proportional $A_{\tau}$. Thus, it becomes of the form $K(t,\tau) A_{\tau}$, where
\begin{align}
\label{kernel_def}
K(t,\tau) = &\left\langle A_{\tau}^*  e^{i\mathcal{L}\tau} \left[i\mathcal{L} - \dot{P}^{b}_{\tau} \right]  \left[1-P^{b}_{\tau}\right]G_{\tau,t}i\mathcal{L}A_{0} \right\rangle \left\langle |A_{\tau}|^{2} \right\rangle^{-1}
\end{align}
This expression further simplifies once one evaluates $\dot P_\tau^b$ explicity.  There holds (see appendix)
\begin{equation}
\dot{P}^{b}_{\tau}F=\frac{\left< A^{*}_{\tau}i\mathcal{L}e^{i\mathcal L\tau}[1-P^b_\tau]F\right>+\left<e^{i\mathcal L\tau}[1-P^b_\tau]Fi\mathcal LA^*_\tau\right>}{\left< |A_{\tau}|^{2} \right>}A_{0}
\end{equation}
Inserting this into eq.~\eqref{kernel_def}, one obtains
\begin{align}
\label{kernel_def_simplified}
K(t,\tau) = &-\left\langle[i\mathcal L A_{\tau}^*]  e^{i\mathcal{L}\tau} \left[1-P^{b}_{\tau}\right]G_{\tau,t}i\mathcal{L}A_{0} \right\rangle \left\langle |A_{\tau}|^{2} \right\rangle^{-1}
\end{align}
Interestingly, the derivative of the projector vanishes in favor of shifting the Liouville operator in $[i\mathcal L-\dot P_\tau^b]$ to act ``to the left'' (an operation that superficially resembles taking the adjoint with respect to the weighted scalar product).

The definition of $G_{\tau,t}$ allows to write 
\begin{align}
\label{kernel_def_with_k}
K(t,\tau) &= k_{0}(\tau) \nonumber \\
&+ \sum_{n=1}^{\infty}{ \int_{\tau}^{t}dt_{1}...\int_{\tau}^{t_{n-1}}dt_{n} k_{n}(\tau,t_{1},...,t_{n})}
\end{align}
The objects $k_{n}$ are functions of the instantaneous correlations of $A$ and its first $n+2$ time-derivatives. They are defined as
\begin{align}
\label{k_proj}
k_{n}(\tau, t_{1},...,t_{n}) =  
&\left\langle A_{0}^* P^{\rm b}_{\tau} \left[i\mathcal{L} - \dot{P^{\rm b}_{\tau}} \right]\left[1-P^{\rm b}_{\tau}\right] \right. \nonumber \\ 
& \left. i\mathcal{L}\left[1-P^{\rm b}_{t_{n}}\right]...i\mathcal{L}\left[1-P^{\rm b}_{t_{1}}\right]i\mathcal{L}A_{0} \right\rangle \left\langle |A_{0}|^{2} \right\rangle^{-1}
\end{align}
Combining these terms, we obtain an equation of motion for $A_{t}$ :
\begin{equation}
\label{EOM_A_full}
\frac{dA_{t}}{dt} = \omega_{1}(t)A_{t} + \int_{s}^{t}{d\tau K(t,\tau) A_{\tau}} + \eta_{s}(s,t)
\end{equation}
with
\begin{equation}
\label{defineEta}
\eta_{s}(s,t) = e^{i\mathcal{L}s}\left[1-P^{\rm b}_{s}\right]G_{s,t}i\mathcal{L}A_{0}
\end{equation} 
The structure of this equation resembles the Generalized Langevin Equation (GLE), however, as the system is not in a steady state, the friction kernel $K(t,\tau)$ does not necessarily depend on only $t-\tau$. 

Note that in the stationary Mori-Zwanzig case, eq.~\eqref{EOM_A_full} is covariant under an arbitrary time translation $s\mapsto s+t_0$, $t\mapsto t+t_0$, and the first term in eq.~\eqref{defineEta} guarantees this covariance for $\eta_s(s,t)$. One can then set $s=0$ to recover the term  that is usually identified as a noise \cite{kubo:1966} with 
\begin{equation}\left\langle \eta^{MZ}(t) \right\rangle = 0 
\end{equation}
and
\begin{equation} \left\langle \eta^{MZ}(t)\eta^{MZ}(t') \right\rangle = -K^{MZ}(t-t')\left\langle |A|^{2} \right\rangle_{\rm eq}
\end{equation} 

In the present case, the situation is more complex. From the definition of $P^{\rm b}_{t}$ follows that $\eta_{s}(s,t)$ is perpendicular to $A_{s}$, i.e. $\left\langle A^{*}_{s} \eta_{s}(s,t) \right\rangle = 0$, but not to $A_T$ for $T\neq s$. The orthogonality can be seen by calculating $\left\langle A^{*}_{s} e^{i\mathcal{L}s}\left[1-P^{\rm b}_{s}\right]F \right\rangle$ with $F = G_{s,t}i\mathcal{L}A_{0}$, i.e.
\begin{equation}
\label{eta_perp}
	\left\langle A^{*}_{s} e^{i\mathcal{L}s}\left[1-P^{\rm b}_{s}\right]F \right\rangle = \left\langle A^{*}_{s} F_{s} \right\rangle - \left\langle A^{*}_{s} \frac{\left\langle A^{*}_{s} F_{s}\right\rangle }{\left\langle |A_{s}|^{2}\right\rangle} A_{s} \right\rangle = 0
\end{equation}
 This allows us to write an equation of motion for the two-time auto-correlation function $C(s,t) = \left\langle A^{*}_{s}A_{t} \right\rangle$ by multiplying eq.~(\ref{EOM_A_full}) by $A^{*}_{s}$ and taking the trajectory average defined in (\ref{traj_ave})
\begin{equation}
\label{EOM_C}
\frac{dC(s,t)}{dt} = \omega_{1}(t)C(s,t) + \int_{s}^{t}{d\tau K(t,\tau) C(s,\tau)} 
\end{equation}
This equation, just as in the stationary case of eqn.~(\ref{simpleGLE}), does not contain the fluctuating force $\eta_{s}(s,t)$ anymore, but
its form respects the non-stationarity of the problem. If one shifts the time origin by an amount $t_0$, i.e.~one considers the equation for $C(s+t_0, t+t_0)$, its solution will be a priori different because $K(t,\tau) \neq K(t+t_0, \tau+t_0)$. Moreover, note that it  is not possible to simply Laplace transform eqn.~(\ref{EOM_C}), because the convolution theroem does not apply anymore. 

Eqn.~(\ref{EOM_A_full}) and eqn.~(\ref{EOM_C}) are the central results of our 
work. Remarkably they differ from their stationary counterparts only in the explicit dependence of the drift, the memory kernel and the fluctuating force on one additional time. Apart from this the structure is the same as in equilibrium.

\subsection{Consequences}
\subsubsection{An FDT-like equation}
 
We now derive a relation between the auto-correlation of the fluctuating force and the memory kernel. The "$\eta$-term'' defined in eqn.~(\ref{defineEta}) and the memory kernel as written in eqn.~\eqref{kernel_def_simplified} are related via
\begin{equation}
K(t,\tau) =  -\frac{\left\langle \eta_{\tau}(\tau,t) i\mathcal{L}A^{*}_{\tau}  \right\rangle}{\left\langle |A_{\tau}|^{2}  \right\rangle}
\end{equation}
Since $\left\langle A^{*}_{\tau} \eta_{\tau}(\tau,t) \right\rangle = 0$ (see eqn.(\ref{eta_perp})), one can write $\left\langle  \eta_{\tau}(\tau,t)i\mathcal{L}A_{\tau}^{*} \right\rangle = \left\langle  \eta_{\tau}(\tau,t) e^{i\mathcal{L}\tau}(1-P^{b}_{\tau}) i\mathcal{L}A_{0}^{*} \right\rangle$. Then, by noticing that $G(\tau,\tau)=1$, one has $\eta_{\tau}(\tau,\tau) = e^{i\mathcal{L}\tau}(1-P^{b}_{\tau})i\mathcal{L}A_{0}$, which finally yields
\begin{equation}
\label{FDT_beautiful}
K(t,\tau) = - \frac{\left\langle \eta^{*}_{\tau}(\tau,\tau)  \eta_{\tau}(\tau,t)  \right\rangle}{\left\langle |A_{\tau}|^{2}\right\rangle} 
\end{equation}
This equation is analogous to the fluctuation-dissipation theorem, but it holds for non-stationary processes. (Other fluctuation-dissipation-like theorems have been derived for non-stationary processes, but in a Fokker-Planck picture \cite{calabrese:2005,verley:2011}). Note, however, that the analogy is just in terms of mathematical structure and not in terms of interpretation.

We can go one step further and try to find a similar relation for the correlation function $\left\langle \eta^{*}_{\tau}(\tau,t)  \eta_{\tau}(\tau,t')  \right\rangle$. To do this, we write 
\begin{equation}
	\eta_{\tau}(\tau,t') = \sum_{n=0}^{\infty} \bar{\eta_{\tau}}^{(n)} \frac{(t'-\tau)^{n}}{n!} 
\end{equation}
where
\begin{equation}
	\bar{\eta_{\tau}}^{(n)} = \lim_{t\rightarrow\tau} \frac{\partial^{n} \eta_{\tau}(\tau,t)}{\partial t^{n}}
\end{equation}
Then, to compute $\bar{\eta_{\tau}}^{(n)}$, we first show from the definition of $\eta_{\tau}(\tau,t)$, using $\partial_{\tau}G_{\tau,t} = -i\mathcal{L}(1-P_{\tau}^{b})G_{\tau,t}$, the identity
\begin{equation}
\label{deta_dtau}
	\frac{\partial \eta_{\tau}}{\partial \tau}(t) = K(t,\tau)A_{\tau}
\end{equation}
This relation will become useful later, but we first need to prove the following relation :
\begin{equation}
\label{recursion}
	\frac{\partial^{n}}{\partial\tau^{n}}\left[K(t,\tau) \left\langle |A_{\tau}|\right\rangle^{2} \right] = -\left\langle  \bar{\eta_{\tau}}^{(n)*} \eta_{\tau}(\tau,t) \right\rangle
\end{equation}
First, the case $n=0$ is true and consists in eqn.(\ref{FDT_beautiful}). Let us assume that eqn.(\ref{recursion}) is true for a certain $n$. Thus, we obtain
\begin{align}
\label{order_n+1}
\frac{\partial^{n+1}}{\partial\tau^{n+1}}\left[K(t,\tau) \left\langle |A_{\tau}|\right\rangle^{2} \right] = &-\left\langle  \partial_{\tau} \bar{\eta_{\tau}}^{(n)*} \eta_{\tau}(\tau,t) \right\rangle \nonumber \\
&- \left\langle   \bar{\eta_{\tau}}^{(n)*} \partial_{\tau}\eta_{\tau}(\tau,t) \right\rangle
\end{align}
Since $\bar{\eta_{\tau}}^{(n)} = (1-P_{\tau})\bar{\eta_{\tau}}^{(n)}$, and $\partial_{\tau}\eta_{\tau}(\tau,t)$ is proportional to $A_{\tau}$ (see eqn.(\ref{deta_dtau})), we have  $\left\langle   \bar{\eta_{\tau}}^{(n)*} \partial_{\tau}\eta_{\tau}(\tau,t) \right\rangle = 0$. We now use 
\begin{equation}
\partial_{\tau} \bar{\eta_{\tau}}^{(n)} = \lim_{t\rightarrow\tau} \left[ \partial_{\tau}\partial_{t}^{n}\eta_{\tau}(\tau,t) + \partial_{t}^{n+1}\eta_{\tau}(\tau,t) \right]
\end{equation}
that we can rewrite as
\begin{equation}
\partial_{\tau} \bar{\eta_{\tau}}^{(n)} = \lim_{t\rightarrow\tau}  [\partial_{t}^{n}K(t,\tau)] A_{\tau} + \bar{\eta_{\tau}}^{(n+1)}
\end{equation}
where we used eqn.(\ref{deta_dtau}). By inserting this last equation into (\ref{order_n+1}), and using again $\left\langle A_{\tau}^{*}\eta_{\tau}(\tau,t) \right\rangle = 0$, we prove that eqn.(\ref{recursion}) is also true at order $n+1$. thus, one can finally write
\begin{equation}
	\left\langle \eta^{*}_{\tau}(\tau,t)  \eta_{\tau}(\tau,t')  \right\rangle = - \sum_{n=0}^{\infty} \frac{\partial^{n}}{\partial\tau^{n}}\left[K(t,\tau) \left\langle |A_{\tau}|\right\rangle^{2} \right] \frac{(t'-\tau)^{n}}{n!}
\end{equation}

 The meaning of the "noise"-term $\eta_{s}(s,t)$ is still not fully clear at this point. In particular, $\eta_{s}(s,t)$ is perpendicular to $A$ only with respect to $P^{b}_{s}$, i.e.~at the initial time $s$. It might thus be interesting to look at its correlation with $A_{t}$. Let us take the derivative of order $n$ with respect to $t$ of equation (\ref{EOM_A_full}) for $A$, multiply it by $A^{*}_{t}$, and take the trajectory average of the result. Then many terms cancel each other, such that one finally finds the following identity
\begin{equation}
\int_{s}^{t}d\tau\frac{\partial^{n} K(t,\tau)}{\partial t^{n}} \left\langle A_{t}^{*}A_{\tau} \right\rangle + \left\langle A_{t}^{*} \frac{\partial^{n} \eta_{s}(s,t)}{\partial t^{n}} \right\rangle = 0
\end{equation}
which is true for all $n$. The time derivative of this relation for the case $n=0$ yields
\begin{align}
\frac{d\left\langle A_{t}^{*} \eta_{s}(s,t)\right\rangle}{dt} = K(t,t) \left\langle |A_{t}|^{2}\right\rangle &+ \int_{s}^{t} d\tau K(t,\tau) \left\langle \dot{A}_{t}^{*}A_{\tau} \right\rangle \nonumber \\ &+ \int_{s}^{t} d\tau \frac{\partial K}{\partial t}(t,\tau) \left\langle A^{*}_{t}A_{\tau} \right\rangle
\end{align}
In the limit of slow processes, we can neglect the last two terms to find
\begin{align}
\frac{d\left\langle A_{t} \eta_{s}(s,t)\right\rangle}{dt} = \left\langle A^{*}_{t}\ddot{A}_{t} \right\rangle = - \left\langle |\dot{A}_{t}|^{2} \right\rangle
\end{align}
which is always negative.

\subsubsection{Time-evolution of the memory kernel}

The main difference between the equations of motion that we derived here and the well-known GLE for the stationary case lies in the explicit dependence of the Kernel on two times. In order to discuss this dependence further, we define the objects $\tilde{K}_{n}(t,\tau)$ by
\begin{equation}
\tilde{K}_{n}(t,\tau) A \equiv P_{\tau} \left[i\mathcal{L} - \dot{P}_{\tau} \right]\left[1-P_{\tau}\right]G_{\tau,t}(i\mathcal{L})^{n}A
\end{equation}
such that $K(t,\tau) = \tilde{K}_{1}(t,\tau)$. Because of the property $\partial_{t}G_{\tau,t} = G_{\tau,t}i\mathcal{L}(1-P_{t})$, we have 
\begin{equation}
\frac{\partial \tilde{K}_{n}(t,\tau)}{\partial t} = -\omega_{n}(t)K(t,\tau) + \tilde{K}_{n+1}(t,\tau)
\end{equation}
By applying successive time-derivatives to $K(t,\tau)$ we obtain
\begin{equation}
\label{inifnite_diffeq}
\lim_{n\rightarrow\infty} \sum_{j=0}^{n} \frac{\partial^{n-j}}{\partial t^{n-j}} \left[ \omega_{j}(t) K(t,\tau) \right] = 0
\end{equation}
Eqn.~(\ref{inifnite_diffeq}) is a linear differential equation of infinite order for which the coefficients are well-controlled functions, meaning that one knows a priori their global properties in most of the situations. For instance, in most physical many-particle processes, $\omega_{n}(t)$ are bounded and infinitely differentiable functions, which ensures the solution of eqn.~(\ref{inifnite_diffeq}) to be a smooth differentiable function. We do thus not expect to see a discontinuous evolution of the memory kernel in such systems.

\subsubsection{A Taylor expansion of the memory kernel}

As shown in the previous paragraphs, the non-stationary memory kernel $K(t,\tau)$ depends explicitly on the times $t$ and $\tau$, while in the stationary case it depends depend only on $t-\tau$. The kernel that we have derived recovers the $t-\tau$ behavior in the stationary limit. To show this, we come back to the definition of $K(t,\tau)$ from eqn.~(\ref{EOM_A}), i.e.~$K(t,\tau) A_{\tau} = e^{i\mathcal{L}\tau}P_{\tau}^{b} \left[i\mathcal{L} - \dot{P}_{\tau}^{b} \right]\left[1-P_{\tau}^{b}\right]G_{\tau,t}i\mathcal{L}A_{0}$. In the stationary limit, $P_{\tau}^{b}$ is a constant operator (i.e.~it does not depend on $\tau$), such that $P_{\tau}^{b}=P^{b}$, $\dot{P}_{\tau}^{b} = 0$ and
\begin{align}
	G_{\tau,t} &= 1 + \sum_{n=1}^{\infty} \left[ i\mathcal{L} (1-P^{b})\right]^{n} \int_{\tau}^{t} dt_{1} \cdots \int_{\tau}^{t_{n-1}}dt_{n} \nonumber \\
	&= \sum_{n=0}^{\infty} \frac{(t-\tau)^{n}}{n!}\left[ i\mathcal{L} (1-P^{b})\right]^{n}
\end{align}
Thus, we obtain
\begin{equation}
K(t,\tau) = \sum_{n=0}^{\infty}\frac{k_{n}^{\infty}}{n!}(t-\tau)^{n}
\end{equation}
with $k_{n}^{\infty} = \left\langle A_{0}^* P^{b} \left[ i\mathcal{L} (1-P^{b})\right]^{n+1} i\mathcal{L}A_{0} \right\rangle \left\langle |A_{0}|^{2} \right\rangle^{-1}$. This limit proves that the well-known dependence of $K$ in $t-\tau$ is recovered as long as the projector $P_{\tau}^{b}$ is constant.

As pointed out above, in the stationary case the dependence of $K$ on $t-\tau$ only, allows to relate the stationary auto-correlation function $C(s,t) = C(0,t-s)$ and $K(t)$ in the Laplace (or Fourier) space, by  making use of the convolution theorem in eqn.~(\ref{EOM_C}). In the non-stationary case, this is no longer possible, thus we need to find another way to evaluate the memory kernel. To do this, and since the integration in eqn.~(\ref{EOM_C}) runs over $\tau$, we perform a one-dimensional Taylor expansion of $K(t,\tau)$ at fixed $t$ and in the direction of $\tau$, around the point $\tau=t$, i.e.
\begin{equation}
 \label{taylor_kernel}
K(t,\tau) = \sum_{n=0}\frac{1}{n!} \kappa_{n}(t) (\tau-t)^{n}
\end{equation}
where $\kappa_{n}(t) \equiv  \partial_{\tau}^{n} K|_{t=\tau}$. These coefficients can be directly computed from the formal definitions (\ref{kernel_def}) and (\ref{k_proj}) of $K(t,\tau)$ and $k_{n}(\tau,t_{1},\cdots,t_{n})$, in which the projection operators are applied only to objects of the form $i\mathcal{L}^{n}A_{0}$, with $n\in\mathbb{N}$. Therefore,  $\kappa_{n}(t)$ can be expressed only in terms of the functions $\omega_{p}(t)$ defined in eqn.~(\ref{omega_def}), with $p \leq n+2$. Let us describe here an example of computation, e.g. for $\kappa_{1}(t)$. First, we show from eqn.~(\ref{kernel_def}) that $\kappa_{1}(t) = \partial_{\tau}k_{0}(t) - k_{1}(t,t)$. Then, from eqn.~(\ref{k_proj}) we get $k_{0}(\tau) = \omega_{2}(\tau) - \omega_{1}^{2}(\tau) - \dot{\omega_{1}}(\tau)$, which yields $\partial_{\tau}k_{0}(t) = \dot{\omega_{2}}(t) - 2\dot{\omega_{1}}\omega_{1}(t) - \ddot{\omega_{1}}(t)$, and $k_{1}(\tau,t_{1}) = \omega_{3}(\tau) - \omega_{2}(\tau)[\omega_{1}(t_{1}) + \omega_{1}(\tau)] +\omega_{1}(t_{1})[\dot{\omega}_{1}(\tau)-\omega_{1}(\tau)^{2}] - \dot{\omega_{2}}(\tau)$, which gives $k_{1}(t,t) = \omega_{3}(t) - 2\omega_{2}(t)\omega_{1}(t) + \omega_{1}(t)\dot{\omega}_{1}(t)-\omega_{1}(t)^{3} - \dot{\omega_{2}}(t)$. We finally obtain $\kappa_{1}(t) = 2\dot{\omega_{2}}(t) - 3\dot{\omega_{1}}(t)\omega_{1}(t) - \ddot{\omega_{1}}(t) - \omega_{3}(t) + 2\omega_{2}(t)\omega_{1}(t) - \omega_{1}(t)^{3}$. Such a procedure can be applied for any order, with increasing complexity. The first orders are reported in the appendix, in which we also show that the number of terms involved in $\kappa_{n}(t)$ grows exponentially with $n$.

We have thus derived a relation between the Taylor coefficients of the memory kernel and the dynamics of the coarse-grained variable as it can be obtained in a MD simulation. In addition, we have shown that despite the complexity of eqn.~\ref{kernel_def} the functional form of the kernel can be constructed without the need to compute an infinite number of nested integrals. 

From this formalism, one can naturally define a timescale $\mathcal{T}$ associated to the time extent of the memory kernel, which may change as the process evolves. In fact,
\begin{equation}
\label{timescale_estimate}
\mathcal{T}^{2}(t) = \left| \frac{\kappa_{0}(t)}{\kappa_{2}(t)} \right| \simeq \left| \frac{\omega_{2}(t)}{\omega_{4}(t)} \right| =  \left| \frac{\left\langle A^{*}_{t} A^{(2)}_{t} \right\rangle}{\left\langle A^{*}_{t} A^{(4)}_{t} \right\rangle} \right|
\end{equation}
provides useful information about the timescale on which the memory kernel is relevant. This quantity can be easily sampled in MD simulations and then used to test for instance a Markovian approximation on the coarse-grained scale.
Second, if one has a theoretical guess for the functional form of the memory kernel, one can test it by constructing the leading Taylor coefficients. Third, and most important, an accurate sampling of $\omega_n$, actually allows to construct the entire ``generalized GLE''. We now show this in a numerical example.

\subsection*{A numerical example}

To illustrate the use of the method, we carried out MD simulations of a two-dimensional model system, defined by one heavy particle of mass $M$ that interacts with bath particles, each of mass $m$, via a potential $V(r) = V_{0}\exp \left(-r/r_{0}\right)$, where $V_{0}$ and $r_{0}$ define the units of energy and distance, respectively. The bath particles do not interact with each other and we set $M=10^{3}m$. The averaged quantity for which we construct an equation of motion is the x-component of the momentum $p$ of the heavy particle, i.e.\  $C(t) = \left\langle p_{x}(t)p_{x}(0) \right\rangle$. 

We initialize the system by placing the heavy particle in the center of box with a velocity drawn from a Gaussian distribution associated to a certain temperature $k_{B}T=10^{-2}V_{0}$. The particles of the bath are initially distributed homogeneously in the box, except in a circular region of radius $R=30r_{0}$ around the central heavy particle. Their velocities are also picked from a Gaussian distribution associated to the same temperature $T$. The boundary conditions are reflective. The system is initially strongly out-of-equilibrium, and it reaches an equilibrium state after going through a transient phase. Thus, it is a well-suited test case for our method. 
\begin{figure}
\begin{center}
\includegraphics[width=.99\linewidth]{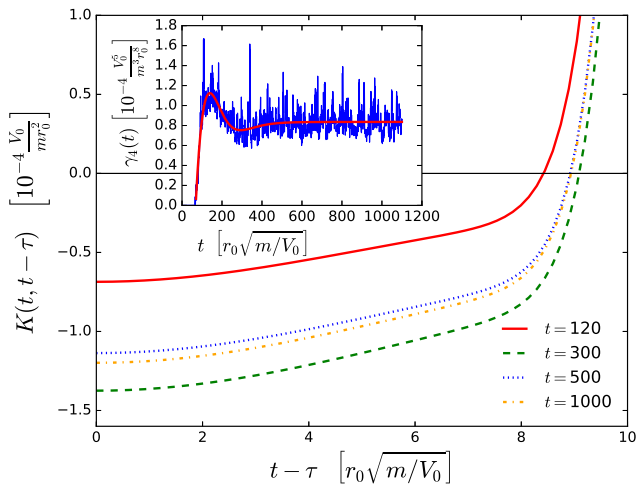}
\end{center}
\caption{Taylor-expansion $K(t,\tau) = \sum_{n}\frac{\kappa_{n}(t)}{n!}(\tau-t)^{n}$ versus $t-\tau$ up to order 18 at various times. The divergence for large value of $t-\tau$ is due to the finite number of terms in the expansion. Note that the form of the kernel depends explicitly on the time $t$ passed since the beginning of the process. The inset shows $\gamma_{4}(t)$, i.e. direct measurement (blue) and an interpolation of it used in practice (red). All $\gamma_{n}(t)$ follow the same global trend.}
\label{gamma_ratios}
\end{figure}

The computation of the local correlations $\left\langle A^{*}(t) A^{(n)}(t) \right\rangle$ is the central operation to perform in order to reconstruct the kernel. As these functions will be obtained by simulation, they can in practice be noisy. To increase the numerical accuracy, we use the following relation (valid for real variables)
\begin{equation}
\label{recursion_correlations}
\left\langle A_{t} A^{(n)}_{t} \right\rangle = \sum_{p=0}^{\lfloor \frac{n}{2} \rfloor} \alpha_{n,p} \frac{d^{n-2p}}{dt^{n-2p}} \left\langle {A^{(p)}_{t}}^{2} \right\rangle
\end{equation}
where $\alpha_{n,0} = 1/2$, $\alpha_{2n,n} = (-1)^{n}$ for all $n$, and the remaining elements are determined by 
\begin{equation}
\alpha_{n,p} = \alpha_{n-1,p} - \alpha_{n-2,p-1}
\end{equation}
Sampling the functions $\gamma_{p}(t) = \left\langle {A^{(p)}_{t}}^{2} \right\rangle$ yields much weaker fluctuations than sampling the functions $\left\langle A_{t} A^{(n)}_{t} \right\rangle$. Note that if we take $A(t) = B^{(m)}(t)$, we obtain
\begin{equation}
\label{recursion_correlations_bis}
\left\langle B^{(m)}_{t} B^{(l)}_{t} \right\rangle = \sum_{p=0}^{\lfloor \frac{l-m}{2} \rfloor} \alpha_{l-m,p} \frac{d^{l-m-2p}}{dt^{l-m-2p}} \left\langle {B^{(p+m)}_{t}}^{2} \right\rangle
\end{equation}

In this example, we sampled $\gamma_{n}(t) = \left\langle \left| d^{n}p_{x}/dt^{n}  \right|^{2} (t) \right\rangle$ in order to compute the functions $\omega_{p}(t)$ and from those the Taylor coefficients $\kappa_{m}(t)$. We used the approximations $\kappa_{2n}(t) = \omega_{2n+2}(t)$ and $\kappa_{2n+1}(t) = 0$, which turn out to be very good in this case. In figure (\ref{gamma_ratios}) we show as an example $\gamma_{4}(t)$ and the function we used to interpolate it. The functional form of $\gamma$ was similar for all orders that we computed (until order 10). We also plot the Taylor expansion of the memory kernel constructed from these measurements, until order 18, as function of $t-\tau$ at various times $t$. We then use use it to solve the equation of motion.
\begin{figure}
\begin{center}
\includegraphics[width=.99\linewidth]{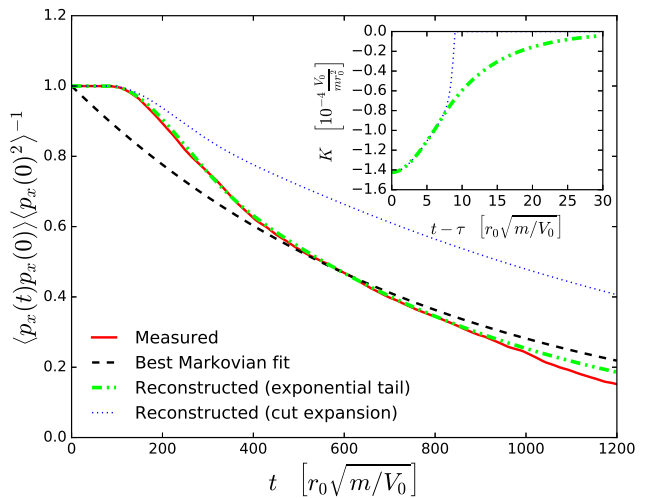}
\end{center}
\caption{$\left\langle p_{x}(t)p_{x}(0) \right\rangle/\left\langle p_{x}(0)^{2} \right\rangle$ from direct simulations (solid line), in Markovian approximation (dashed line), and computed using the kernel constructed by means of the method described in this paper (dotted and dash-dotted, see main text for details). The inset shows $K(t,t-\tau)$ at time $t=20$.}
\label{autocorr}
\end{figure}
Fig.~\ref{autocorr} shows the momentum auto-correlation obtained directly from the MD simulation (solid line), i.e.\ the data extracted from the full ``microscopic dynamics''. The dashed line is the best approximation that one can get if making the assumption that the dynamics of the averaged observable is Markovian. This approximation is commonly used in coarse-graining procedures. It clearly fails here, as it does not capture the short time-plateau. 

The dotted and dash-dotted lines have been obtained by constructing the memory kernel according to eqn.~\ref{taylor_kernel} up to order 18 and then solving the generalized Langevin equation. Both capture the initial plateau very accurately.  

If one approximates the kernel until it changes sign and then simply truncates it (dotted line), the long time behaviour of $\left\langle p_{x}(t)p_{x}(0) \right\rangle$ is not reproduced convincingly, because the expansion diverges. This can be fixed by extrapolating a tail on $K$ for large values of $t-\tau$. From the equilibrium dynamics, it is well known that the VACF and its memory kernel exhibit a $t^{-d/2}$ long-time tail, where $d$ is the dimension \cite{pomeau:1975,corngold:1972}. This would suggest extrapolation by an algebraic tail proportional to $(t-\tau)^{-1}$. However, the transient VACF starting from the non-equilibrium configuration does not exhibit such a long-time tail. If we fit an exponential decay, as is indicated by the large $n$ behaviour of the observed $\kappa_n$, the coarse-grained description (dash-dotted line)  accurately captures the features of the directly computed correlation function. We have thus succeeded in constructing a coarse-grained description of this non-equilibrium model system.  

\subsection{Summary}

~\newline

We have introduced a time-dependent projection operator that is of practical use, if one whishes to study non-equilibrium trajectory averages of phase space variables. We showed that, in the case of non-stationary dynamics, the equation of motion for the trajectory averages, eqn.~(\ref{EOM_A_full}), resembles the Generalized Langevin Equation. The only difference is an explicit dependence on an additional time in the drift term, the memory kernel and the fluctuating force. For all practical cases of application, the memory kernel is a smooth function in the additional time.  We also derived an equation of motion for the auto-correlation function of the observable, eqn.~(\ref{EOM_C}). Remarkably, as in the stationary Mori-Zwanzig case, this equation does not contain noise.

We also showed how to systematically construct the memory kernel of a non-stationary GLE using as input data from experiments or MD simulations of the underlying microscopic dynamics. We thus provide a general strategy to develop coarse-graining procedures in classical atomistic computer simulations. In particular, we Taylor-expand the kernel and express its coefficients in terms instantaneous correlation functions of the variable of interest with its consecutive time-derivatives. If one can accurately measure time-derivatives up to order $n$, one can Taylor-expand until order $2(n-1)$. This allows to infer how long the system keeps track of its history, and can thus be used to test approximations that are often made in simulations on the coarse-grained scale (as e.g.\ the assumption of Markovian dynamics). If those approximations fail, the method can be used to construct appropriate equations of motion.

\section{Acknowledgements}
We thank T.\ Franosch, A.\ Kuhnhold, M.\ Dolgushev and G.\ Amati for useful discussions. This project has been financially supported by the National Research Fund Luxembourg (FNR) within the AFR-PhD programme. Computer simulations presented in this paper were carried out using the HPC facility of the University of Luxembourg.

\newpage

\subsection*{Appendix}
 
\subsection*{Link to microscopic dynamics and a semi-analytic example}
Here, we show an example of how to link the functions $\omega_{n}(t)$ to the microscopic dynamics of a system in practice, and how to use them to reconstruct the memory kernel. We focus on the case of Hamiltonian dynamics in a microscopic many-particle system, i.e.
\begin{equation}
\mathcal{H} = \sum_{\alpha}\frac{p_{\alpha}^{2}}{2m_{\alpha}} + \sum_{\alpha,\alpha'}V(q_{\alpha},q_{\alpha'})
\end{equation}
where $q$, $p$ denote position and momentum, respectively. To compute $\omega_{n}(t) = \left\langle A_{t}A^{(n)}_{t} \right\rangle/\left\langle A_{t}^{2} \right\rangle$, the obvious first step is to calculate $A^{(n)} = (i\mathcal{L})^{n}A$, with 
\begin{equation}
i\mathcal{L} = \sum_{\alpha} \frac{p_{\alpha}}{m_{\alpha}}\frac{\partial}{\partial q_{\alpha}} - \sum_{\alpha,\alpha'} \frac{\partial V_{\alpha,\alpha'}}{\partial q_{\alpha}}\frac{\partial}{\partial p_{\alpha}}
\end{equation}
with $V_{\alpha,\alpha'} = V(q_{\alpha},q_{\alpha'})$. Of course, such a calculation becomes very lengthy and untractable for a non-specified generic variable. Therefore, we will do it on a relevant example, to show how such a study can be carried out in a practical case. 

\begin{figure}
	\begin{center}
		\includegraphics[width=\linewidth]{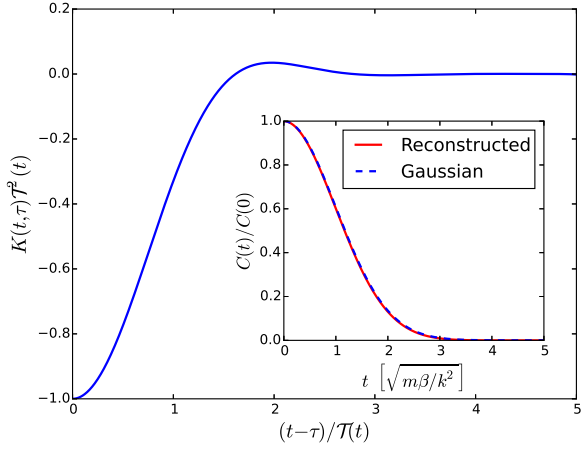}
	\end{center}
	\caption{$K(t,\tau)\mathcal{T}^{2}(t)$ vs. $(t-\tau)/\mathcal{T}(t)$, from the analytic result for $\omega_{n}(t)$. The Taylor expansion is computed until order 80. The inset shows $f_{n}$ as a function of $n$.}
	\label{analytic_kernel}
\end{figure}

We choose a variable $A$ which depends only on the positions and is of the form
\begin{equation}
A \equiv \sum_{\alpha} c_{\alpha}f(q_{\alpha})
\end{equation} 
For this type of variables, one can show 
\begin{equation}
\frac{d^{n}A}{dt^{n}} = \sum_{\alpha}c_{\alpha}  \frac{\partial^{n} f}{\partial q_{\alpha}^{n}}  \frac{p_{\alpha}^{n}}{m_{\alpha}^{n}} + \sum_{\alpha_{1},\cdots,\alpha_{n}}c_{\alpha_{1}}\mathcal{V}_{\alpha_{1},\cdots,\alpha_{n}}
\end{equation}
where $\mathcal{V}_{\alpha_{1},\cdots,\alpha_{n}}$ is a object involving various derivatives of the potentials with respect to the positions $q_{\alpha_{1},\cdots,\alpha_{n}}$, as well as the momenta $p_{\alpha_{1},\cdots,\alpha_{n}}$ and derivatives of $f(q_{\alpha_{1}})$. Thus, we obtain
\begin{align}
\label{micro_AAn}
\left\langle A_{t}A^{(n)}_{t} \right\rangle = &\sum_{\alpha,\gamma}c_{\alpha}c_{\gamma}  \left\langle \frac{\partial^{n} f}{\partial q_{\alpha}^{n}} f_{\gamma}\right\rangle  \frac{\left\langle p_{\alpha}^{n}\right\rangle}{m_{\alpha}^{n}} \nonumber \\
&+ \sum_{\substack{\alpha_{1},\cdots,\alpha_{n} \\ \gamma } } c_{\alpha_{1}}c_{\gamma} \left\langle \mathcal{V}_{\alpha_{1},\cdots,\alpha_{n}} f_{\gamma} \right\rangle
\end{align}
where $f_{\gamma} = f(q_{\gamma})$. As a specific case, we now consider a system of identical particles $(m_{\alpha} = m_{\alpha'} = m, \left\langle p_{\alpha}^{n}\right\rangle = \left\langle p_{\alpha'}^{n}\right\rangle = \left\langle p^{n}\right\rangle)$, and we compute the density fluctuations, i.e.
\begin{equation}
A_{k}(q_{\alpha}) \propto \sum_{\alpha} e^{ikq_{\alpha}}
\end{equation}
We rewrite now equation (\ref{micro_AAn}) as
\begin{align}
\label{micro_AAn_flu}
\left\langle A_{k}(t)A_{k}^{(n)}(t) \right\rangle = &\frac{\left\langle p^{n}(t)\right\rangle}{m^{n}} (ik)^{n}\sum_{\alpha,\gamma}  \left\langle f_{\alpha} f_{\gamma}\right\rangle  \nonumber \\
&+ \sum_{\substack{\alpha_{1},\cdots,\alpha_{n} \\ \gamma } }  \left\langle \mathcal{V}_{\alpha_{1},\cdots,\alpha_{n}} f_{\gamma} \right\rangle
\end{align}
In a high-temperature regime (or for weakly interacting systems), the second term can be neglected with respect to the first one. Thus, we have
\begin{equation}
\omega_{n}(t) = \left\langle p^{n}(t)\right\rangle (ik/m)^{n}
\end{equation}
Finally, we assume that the phase-space distribution remains symmetric with respect to momenta and is of the shape
\begin{equation}
\label{assume_can}
\rho(\mathbf{q}^{N},\mathbf{p}^{N},t) \propto e^{-\beta(t)\sum_{\alpha}p_{\alpha}^{2}/2m} \rho_{q}(\mathbf{q}^{N},t)
\end{equation} 
This approximation consists in assuming a slow relaxation of the system towards equilibrium. We obtain then
\begin{align}
\omega_{2n+1}(t) &= 0 \\
\omega_{2n}(t) &= \frac{(-2)^{n}}{\sqrt{\pi}}\Gamma\left(n+\frac{1}{2}\right)\left(\frac{k^{2}}{m\beta(t)} \right)^{n}
\end{align}
The assumption (\ref{assume_can}) is valid only if the system evolves slowly. Thus, we assume that the derivatives of $\omega_{n}(t)$ involved in the Taylor coefficients of the memory kernel are also negligible, yielding  $\kappa_{0}(t) = \omega_{2}(t)$, $\kappa_{2}(t) = \omega_{4}(t) - \omega_{2}^{2}(t)$, $\kappa_{4}(t)  = \omega_{6}(t) - 2\omega_{4}(t)\omega_{2}(t) + \omega_{2}^{3}(t)$, ... Because of the scaling of $\omega_{2n}$ with $n$, we can write 
\begin{equation}
\kappa_{2n}  = f_{2n} \left(\frac{k^{2}}{m\beta(t)} \right)^{n+1}
\end{equation} 
where we calculate the coefficients $f_{2n}$ analytically from our formalism for the first orders. As an example we have $\kappa_{2}=\omega_{4}-\omega_{2}^2 = (2/3)\omega_{4}$, i.e. $f_{2}=2/3$. The Taylor expansion becomes then
\begin{equation}
K(t,\tau) = \mathcal{T}(t)^{-2} \sum_{n}^{\infty} \frac{f_{2n}}{(2n)!}  \left[\frac{t-\tau}{\mathcal{T}(t)}\right]^{2n}
\end{equation}
where $\mathcal{T}(t)^{2} = m\beta(t)/k^2$. This sum can be numerically computed until very large orders without effort. We show in figure (\ref{analytic_kernel}) the resulting reconstructed kernel (until order 80), as well as the solution of equation (\ref{EOM_C}) for $C(t)$ using the latter kernel with a constant $\mathcal{T}$. The agreement of the reconstructed correlation function with well-known result $C(t) = C_{0} \exp \left[-t^{2}k^{2}/2m\beta \right]$ at high temperature \cite{hansen:1990, wallace:2016} is perfect. Of course, this result holds only within the assumptions made for this specific case, but the calculation shows that it may be possible to retrieve useful information about the functions $\omega_{n}(t)$ from the microscopic dynamics, and hence to partially infer the friction kernel. One can attempt to apply this sort of method for other types of processes and variables.

\subsection*{Computation of the Taylor coefficients in practice}

Here, we show how to compute $\kappa_{n}$ practically. We first define the quantity 
\begin{widetext}
	\begin{align}
	I_{p,n,l}^{m_{0},m_{p+1},\cdots,m_{n}} \equiv \int_{\tau}^{t}dt_{1}\cdots\int_{\tau}^{t_{p-1}}dt_{p} \frac{\partial^{l}k_{n}}{\partial^{m_{0}}\tau\partial^{m_{p+1}}t_{p+1}\cdots\partial^{m_{n}}t_{n}}(\tau,t_{1},\cdots,t_{p},\tau,\cdots,\tau)
	\end{align}
\end{widetext}
The index l stands here for the order of derivation and is such that $m_{0} +\sum_{i=p+1}^{n}m_{i} = l$. Now, we show that the derivative of this object with respect to $\tau$ obeys to the identity 
\begin{align}
\frac{\partial}{\partial\tau}I_{p,n,l}^{m_{0},m_{p+1},\cdots,m_{n}} = &(\delta_{p,0}-1)I_{p-1,n,l}^{m_{0},0,m_{p+1},\cdots,m_{n}} \nonumber \\
&+ I_{p,n,l+1}^{m_{0}+1,m_{p+1},\cdots,m_{n}} \nonumber \\
&+ \sum_{i=p+1}^{n}I_{p,n,l+1}^{m_{0},m_{p+1},\cdots,m_{i}+1,\cdots,m_{n}}
\end{align}
where $\delta$ is the Kronecker symbol. The way this relation is used to find the Taylor coefficients is quite easy then. The first thing is to write the kernel $K(t,\tau)$ as a sum of integrals $I$ :
\begin{equation}
K(t,\tau) = \sum_{n=0}^{\infty}I_{n,n,0}^{0}
\end{equation}
To obtain the Taylor coeffient of order m, we apply then recursively the identity (2) to the $m + 1$ first terms of the sum, and then we keep only the terms with $p = 0$ (corresponding to the limit $\tau \rightarrow t$). As an example, let us compute $\kappa_{1}$.

\begin{align}
\frac{\partial K}{\partial \tau}(t,\tau) &= \boxed{\frac{\partial}{\partial\tau}I_{0,0,0}^{0}} + \boxed{\frac{\partial}{\partial\tau}I_{1,1,0}^{0}} \nonumber \\
&= \boxed{I_{0,0,1}^{1}} + \boxed{-I_{0,1,0}^{0,0} + I_{1,1,1}^{1,0}} \nonumber \\
&= \frac{\partial k_{0}}{\partial\tau}(\tau) - k_{1}(\tau,\tau) + \int_{\tau}^{t}dt_{1}\frac{\partial k_{1}}{\partial\tau}(\tau,t_{1}) \nonumber  \\
& \xrightarrow[\tau \rightarrow t]{} I_{0,0,1}^{1} - I_{0,1,0}^{0,0} = \frac{\partial k_{0}}{\partial\tau}(t) - k_{1}(t,t) = \kappa_{1}(t)
\end{align}
Once one has the expression of $\kappa_{n}$ as a function of $k_{m}$, one can insert the expression of these function in terms of $\omega_{p}$ from eq. (\ref{k_proj}). We show here as a an example $k_{2}(\tau,t_{1},t_{2})$ :
\begin{align}
&k_{2}(\tau,t_{1},t_{2}) = \dot{\omega_{1}}(\tau)\left[\omega_{2}(t_{2})-\omega_{1}(t_{1})\omega_{1}(t_{2})\right]+\omega_{1}(t_{1})\dot{\omega_{2}}(\tau) \nonumber \\
& -\dot{\omega_{3}}(\tau) + \omega_{1}(t_{1})\left[\omega_{1}(\tau)\omega_{2}(\tau)-\omega_{1}(\tau)^2\omega_{1}(t_{2}) \right. \nonumber \\
&\left. +\omega_{2}(\tau)\omega_{1}(t_{2})-\omega_{3}(\tau)\right]  +\omega_{1}(\tau)^2\omega_{2}(t_{2})-\omega_{2}(\tau)\omega_{2}(t_{2}) \nonumber \\
&-\omega_{1}(\tau)\omega_{3}(\tau) +\omega_{4}(\tau)
\end{align}
One finally can express the sum as a function of $\gamma_{q}$ by using eq. (\ref{recursion_correlations}). Again, as an example, we have for $\omega_{2}(t)$ :
\begin{equation}
\omega_{2}(t) = \left(\frac{1}{2}\dot{\gamma_{0}}(t) - \gamma{1}(t) \right)\gamma_{0}(t)^{-1}
\end{equation}
We show here the example of $\kappa_{2}$ as a function of $\gamma_{q}$
\begin{align}
\kappa_{2}(t)  =& \frac{1}{\gamma_{0}(t)^{4}}\left[-\frac{9}{8}\dot{\gamma_{0}}(t)^{2}\gamma_{0}(t)\ddot{\gamma_{0}}(t)+\frac{1}{4}\dot{\gamma_{0}}(t)\gamma_{0}(t)^{2}\dot{\ddot{\gamma_{0}}}(t)\right. \nonumber \\
&+ \left. 2\ddot{\gamma_{0}}(t)\gamma_{0}(t)^{2}\gamma_{1}(t)-\frac{9}{4}\dot{\gamma_{0}}(t)^{2}\gamma_{0}(t)\gamma_{1}(t)\right. \nonumber \\
&+ \left. \frac{1}{2}\dot{\gamma_{0}}(t)\gamma_{0}(t)^{2}\dot{\gamma_{1}}(t) - \frac{1}{2}\gamma_{0}(t)^{3}\ddot{\gamma_{1}}(t)+\frac{15}{16}\dot{\gamma_{0}}(t)^{4}\right. \nonumber \\
&+ \left.\gamma_{0}(t)^{3}\gamma_{2}(t)-\gamma_{0}(t)^{2}\gamma_{1}(t)^{2}\right]
\end{align}
In general, one can express $\kappa_{n}(t)$ as
\begin{equation}
\kappa_{n}(t) = \frac{1}{\gamma_{0}(t)^{n+2}}\sum_{k} \alpha_{k} \prod_{i,j} \left( \frac{d^{j}\gamma_{i}}{dt^{j}} (t) \right)^{p_{i,j}^{(k)}}
\end{equation}
The sum runs over all possible terms for which the dimension is the same on both sides of the equal sign. The coefficients $\alpha_{k}$ are calculated by the method presented in the previous lines. For each term of the sum, one must thus have $\sum_{i,j} p_{i,j}^{(k)} = n+2$ and $\sum_{i,j} (2i+j) p_{i,j}^{(k)} = n+2$. One can show that the number of terms in the sum grows roughly exponentially with $n$.

\vspace{1.5ex}
	
\subsection*{Action of $\dot{P}_{\tau}^{b}$}
		
Here, we show how the operator $\dot{P}_{\tau}^{b}$ acts on an arbitrary dynamical variable $F$. One has
\begin{equation}
  \dot P^b_\tau F_0=A_0\frac{d}{d\tau}\frac{\left< A_\tau^*F_\tau\right>}
  {\left<|A_\tau|^2\right>}
\end{equation}
which can be written out as
\begin{widetext}
\begin{equation}
  \dot P^b_\tau F_0=A_0\frac{
  \left<A^*_\tau i\mathcal LF_\tau\right>+\left<F_\tau i\mathcal LA^*_\tau\right>
  -(\left<A^*_\tau F_\tau\right>/\left<|A_\tau|^2\right>)
  (\left<A^*_\tau i\mathcal LA_\tau\right>+\left<A_\tau i\mathcal LA^*_\tau\right>)
  }{\left<|A_\tau|^2\right>}
  =A_0\frac{
  \left<A_\tau^*i\mathcal LB_\tau\right>+\left<B_\tau i\mathcal LA_\tau^*\right>
  }{\left<|A_\tau|^2\right>}
\end{equation}
\end{widetext}
where we have defined $B_\tau=F_\tau-A_\tau\left<A_\tau^*F_\tau\right>/\left<|A_\tau|^2\right>$. But the last term in $B_\tau$ is simply $e^{i\mathcal L\tau}P_\tau^bF_0$, and thus we obtain
\begin{equation}
  \dot P_\tau^bF_0=A_0\frac{
  \left<[i\mathcal LA_\tau^*]e^{i\mathcal L\tau}(1-P_\tau^b)F_0\right>
  +\left<A_\tau^*i\mathcal Le^{i\mathcal L\tau}(1-P_\tau^b)F_0\right>
  }{\left<|A_\tau|^2\right>}
\end{equation}

\end{document}